\documentstyle[psfig,prl,aps]{revtex} 
\psfigurepath{figs}
\begin{document}
\title{Stressed backbone and elasticity of random central-force systems}
\author{C.~Moukarzel
\footnote{Present Address: Instituto de Fisica, Universidade Federal
Fluminense, Niteroi RJ, Brazil. \\ e-mail: cristian@if.uff.br}}
\address{
H\"ochstleistungsrechenzentrum, Forschungszentrum J\"ulich,\\
D-52425 J\"ulich, Germany.}
\author{P.~M.~Duxbury}
\address{ Dept. of Physics/Ast. and Center for Fundamental Materials
Research,\\ Michigan State University, East Lansing, MI 48824, USA.
}
\maketitle
\begin{abstract}
We use a new algorithm to find the stress-carrying backbone of
``generic''  site-diluted triangular lattices of up to $10^6$ sites.
Generic lattices can be made by randomly displacing the sites of a
regular lattice (see Fig. 1).  The percolation threshold is $p_c=0.6975
\pm 0.0003$, the correlation length exponent $\nu =1.16\pm 0.03$ and the
fractal dimension of the backbone $D_b=1.78 \pm 0.02$.  The number of
``critical bonds'' (if you remove them rigidity is lost) on the backbone
scales as $L^{x}$, with $x=0.85 \pm 0.05$.  The Young's modulus is also
calculated.
\end{abstract}
\twocolumn
The forces between atoms can often be divided into two classes ``central
forces'' and ``angular forces'' (e.g. covalent bonds).  In engineering,
structures composed of bars connected at nodes (e.g.  some bridges), get
their rigidity primarily from the tensile and compressive stiffness of
the bars (these are central-force terms).  Structures of this sort are
called ``trusses'', while those in which the angle forces (or
beam-bending) are important are called ``frames''.  It is simple to see
that systems which are dominated by angle forces support an applied
stress as long as they are simply connected.  In contrast, systems with
only central forces require higher order connectivity, the simplest
rigid structure being a triangle.  In many applications; for example in
granular media$^1$, glasses$^2$, gels$^{3,4}$ and in engineering design,
the disorder in a central-force structure is important and must be
considered.
The stress-bearing paths of central-force systems have been primarily
studied by brute-force solution of the force equations$^{5-8}$.
Although useful and important, this method is slow and subject to
 roundoff errors for large structures.  An efficient method for relating
the {\it  connectivity} of a central-force structure to its ability to
carry stress is an important and, in general unsolved, problem.  One
exception to this is two-dimensional random lattices, for which exact
conditions$^{9-12}$  relating connectivity to ``rigidity'' have existed
for over a decade.  However, till recently$^{13}$ there has been no
efficient implementation of these conditions, and their associated
algorithms in either physics or engineering.   This paper and the
preceding paper by Jacobs and Thorpe (JT)$^{13}$ describe the first
implementations of these ideas.   We use our algorithm, to calculate the
stressed backbone, and in combination with an iterative solver, to find
the elastic properties of these backbones.  We also identify the {\it
critical (red) bonds} as those whose removal would lead to loss of
rigidity, and study their scaling properties.  For reasons outlined
below these methods apply to randomly displaced (or ``generic'' - see
Fig. 1a) central-force lattices.
\begin{figure}[] \vbox{ 
\centerline{ \psfig{figure=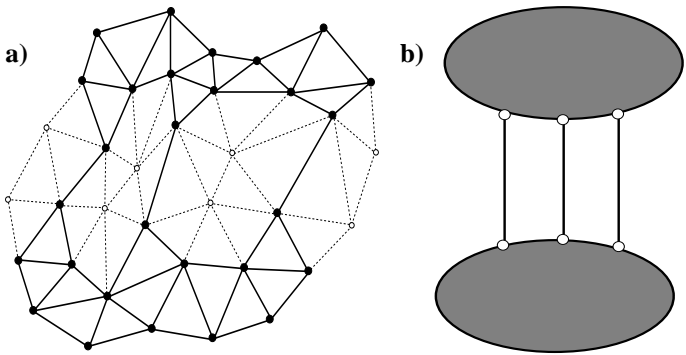,width=9cm,angle=270} }
\centerline{}
\caption{ A configuration that is unstable to shear on a regular lattice, but is
stable on a displaced lattice (dotted lines indicate absent bonds). {\bf
a)} The configuration in the ``bar-joint'' representation ($28$ joints
and $53$ bars).  {\bf b)} The configuration in the ``body-bar''
representation ($2$ bodies and $3$ bars).}
\label{fig:one}
} \end{figure}
Our ability to determine, from connectivity information alone, whether a
central-force structure contains a stress carrying path is based on
Laman's theorem$^9$.
\\
{\bf Laman's theorem} {\it A random lattice (see below for a precise
definition) consisting of $N$ nodes and $B$ bonds so that $2N-B=3$ is
rigid {\it if and only if} there is no subset of the lattice, consisting
of $n$ nodes connected by $b$ bonds, for which $2n-b \leq 3$ is
violated}.
\\
This is the ``bar-joint'' statement of Laman's theorem.  The origin of
the expression $2n-b=3$ is easy to understand.  Each node (joint) in two
dimensions has two degrees of freedom (two translations), and each bond
(bar) is a constraint (for example in Fig. 1a, $n=28$, $b=53$). In the
expression $2n-b=3$, the 3 is there because in two dimensions a rigid
body (in this case the whole lattice or cluster) has 3 degrees of
freedom (two translations and a rotation).  $2n-b=3$ is the two
dimensional version of a general constraint counting argument introduced
by Maxwell.
\begin{figure}[] \vbox{ 
\centerline{ \psfig{figure=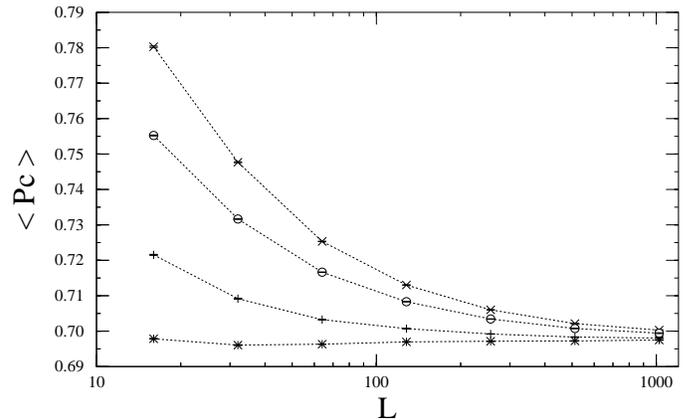,width=9cm,angle=090} }
\centerline{}
\caption{ 
The rigidity threshold as a function of sample size.  AS with periodic
boundary conditions (*), AS with open boundary conditions(+), IS with
open boundary conditions ($\times$) IS with periodic boundary
 conditions($\circ$).  The lattice sizes (L) (number of configurations)
used are as follows; $16(2 \times 10^5)$, $32(10^5)$, $64(8 \times
10^4)$, $128(2 \times 10^4)$, $256(1.2 \times 10^4)$, $512(2 \times
10^3)$, $1024(2 \times 10^2)$.
}
\label{fig:two}
} \end{figure}
However the new feature here is that constraint counting is exact in two
dimensions provided it is implemented at all length scales
(Unfortunately this result does not extend to three dimsensions, where
counterexamples$^{12}$ to the three-dimensional extension of this
argument, $3n-b=6$, are known to exist). However, even in 2-d a naive
algorithm must check all subclusters of a set of $N$ nodes and so is not
polynomial complete.  However Laman's theorem may be implemented by
using the ``bipartite matching'' algorithm from graph theory$^{12}$,
which, when refined as described below, scales as $N^{1.15}$ for finding
the stressed backbone at the rigidity percolation point.
\\
Our implementation of Laman's theorem is a cluster labelling
algorithm$^{21}$.  Although we do site percolation, where $p$ is the
probability that a site is occupied, the algorithm works by testing a
newly added bond against the configuration of rigid clusters already on
the lattice.  For a given $p$, we find the site configuration, and from
it the configuration of present bonds.  Then we start with an empty
lattice and add the present bonds one at a time. Each rigid cluster is a
``body'' with 3 degrees of freedom, so we must generalise the statement
$2n-b=3$ of the original lattice to  $3n_{bod} - b = 3$, where $n_{bod}$
is the number of bodies (or rigid clusters) in a configuration.  For
 example the configuration of Fig. 1a, has two bodies and 3 bars (see
Fig. 1b). A key component of the algorithm is the realisation by
Hendrickson$^{12}$ that it is easy to determine whether a bar (bond) is
 {\it redundant} with respect to the bonds that are already in the
lattice.
\begin{figure}[] \vbox{ 
\centerline{ \psfig{figure=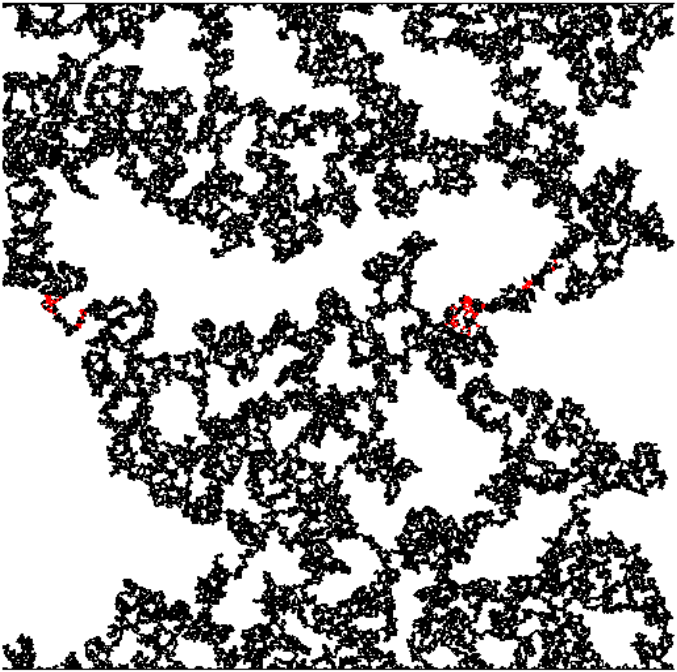,width=9cm,angle=000,clip=58 1148 517
606} }
\centerline{}
\caption{ The stressed (AS) backbone on a lattice of size $L=1024$ with
open boundaries.
 }
\label{fig:three}
} \end{figure}
If the bonds of the lattice are replaced by Hooke's springs, a redundant
bond leads to internal stresses in the spring lattice.  A redundant bond
causes a violation of the condition $3n_{bod}-b=3$ by adding an extra
spring to the lattice.  The algorithm checks to see if $3n_{bod}-b=3$ is
satisfied by using ``bipartite matching''$^{12}$ or ``the pebble
game''$^{12,13}$  to see if each of the bodies' degrees of freedom can
be ``matched'' to the bonds of the configuration (note that JT use the
``pebble game'' in the original bar-joint representation).
\\
If an extra or redundant spring is added to a cluster that is {\it
already rigid} (i.e. already satisfies $3n_{bod}-b=3$), the matching
 algorithm identifies the bonds which become internally stressed when
 the extra spring is added.   We then give these internally stressed
bonds the same cluster label.   In this way we find ``internally
stressed (IS) clusters''.  Finally, an applied tensile stress can be
mimiced by adding two rigid beams to the two opposite sides of the
lattice, and then by adding a fictitious bond between the rigid beams.
In this way, we determine when the lattice can support an applied
tensile stress.  Full details of the algorithm will appear
elsewhere$^{21}$.  In JT, the bond probability is fixed.  Our algorithm
is complimentary to theirs as we add bonds one at a time, so we find the
percolation concentration {\it exactly} for each sample. We chose this
method not only because it is very efficient (comparable to JT), but
also so that we can find the exact backbone for each sample, and hence
the ``critical (or red) bonds'' of the backbone (see below).
\\
There are several ways to define the onset of stress transmission
through a lattice. The two which are most physically appealing are:
\\
\begin{enumerate}
\item The point at which {\it an applied stress (AS)} is transmitted
across the lattice and;
\item The point at which {\it internally stressed regions (IS)} connect
together to form stressed clusters of macroscopic size.
\end{enumerate}

Both of these definitions have simple representations in terms of a
lattice of Hooke's springs.  The first (AS) corresponds to a random
Hooke's spring lattice to which, for example, a tensile stress is
applied, while the second (IS) corresponds to the internal stresses in a
random Hooke's spring lattice with random natural lengths.  We study the
stressed backbone of these lattices as a function of site dilution.  We
also tested the effect of boundary conditions on these two definitions
of rigidity percolation, because a local change in rigidity (e.g. by
adding a bond) can be transmitted over long distances  so boundaries
might be more important in this problem than in connectivity
percolation.  However, we find that in the large-lattice limit both the
AS and IS percolation definitions lead to the same, boundary condition
independent, threshold.  This behavior is presented in Fig. 2, from
which we find that $p_c = 0.6975 \pm 0.0003$.  On regular lattices,
previous work$^{15}$ using direct solution of the force equations lead
to estimates close to $p_c = 0.713$ for samples of up to size $L=75$.
As can be seen from Fig. 2, this is consistent with the result on random
lattices, although at that lattice size, there is still considerable
dependence on boundary conditions. However, in general there is no
reason to believe that the percolation threshold on random lattices
should be the same as that on regular lattices.  This difference is
illustrated by the configuration of Fig. 1b.  On a regular lattice that
configuration is not rigid to shear, but if the lattice sites are
displaced, it becomes rigid. That is because on a regular lattice, the
three bars are parallel, so these constraints are ``degenerate''.   Thus
for that configuration, the random lattice is more rigid than the
regular lattice.
\begin{figure}[] \vbox{ 
\centerline{ \psfig{figure=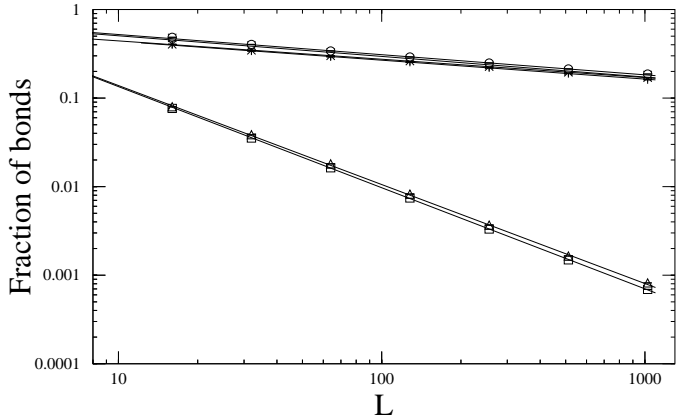,width=9cm,angle=090} }
\centerline{}
\caption{ 
{\bf a)} Finite-size-scaling plot of the normalised number of bonds
$N_B/L^2$ on the stressed backbone at the percolation point for: AS with
periodic b.c.'s (*); AS with open b.c.'s (+); IS with periodic b.c.'s
(X) and ; IS with open b.c.'s (O). {\bf b)}  Finite-size-scaling plot of
the normalised number of ``red'' bonds on the backbone, $N_R/L^2$ (AS
with periodic b.c's).  The lattice sizes used and number of
configurations were the same as for Fig. 2.
 }
\label{fig:four}
} \end{figure}
However it is easy to construct configurations which are more rigid on a
regular lattice (e.g. a sequence of aligned bonds forming a ``guy''
wire), so it is unclear as to whether displaced lattices have a lower or
higher $p_c$ than regular lattices.
\\
From the variation in the percolation concentration $\Delta p_c \sim
L^{-1/\nu}$, we are able to find the correlation length exponent.  We
did this for three ways of defining $\Delta p_c$, namely:
\\
 a) $(p_c(L) - p_c(\infty))$; \\
 b) $(p_c^{open} - p_c^{periodic})$ and; \\
 c) $\surd(<p_c^2>-<p_c>^2)$, \\
and for several types of boundary conditions in each case.  From these
extensive calculations, we find $\nu = 1.16 \pm 0.03$.  Although it is
not the main focus of this paper, we note that in JT, it is claimed that
the infinite cluster, $P_{\infty}$ (which includes internally stressed
bonds (stressed backbone), {\it and} unstressed bonds which satisfy
$2n-b=3$) has a fractal dimension around $D_f \sim 1.86$.
\\
If we assume a second order behavior in $P_{\infty}$ , we find a similar
fractal dimension.  However, a mean field theory$^4$ suggests that the
rigidity transition is first order (so $D_f=2$ in 2-d), and similarly on
Bethe lattices the rigidity transition is first order$^{22}$.  Thus we
have also tested the possibility of a first order transition$^{23}$  in
$P_{\infty}$, and find that the data is consistent with a weakly
first-order transition, with the first-order jump $\Delta P_{\infty}\sim
 0.085$ at $p_c$.  However, even larger lattices (up to of order
$L=10,000$) are needed to determine convincingly whether, in 2-d,
 $P_{\infty}$ is first order.
\\
An example of a stressed backbone at the percolation point is presented
in Fig. 3.  We measured the number of bonds on backbones such as that
shown in Fig. 3, and the results of a scaling plot are presented in Fig.
4.  From this figure, we find $D_b = 1.78 \pm 0.02$.  This backbone
dimension is different than that for connectivity percolation where the
backbone dimension is $1.62 \pm 0.01$, and it is also considerably
larger than that of the stressed backbone of regular triangular lattices
$1.64 \pm 0.05$ found by direct solution$^8$ on small lattices (up to
$L=80$).  The latter discrepancy could be due to a fundamental
difference between the random and regular lattices, but it also could be
due to imprecise estimates of the percolation concentration in previous
work due to finite size effects (see Fig. 2).  At the percolation point,
there are a set of bonds whose removal leads to loss of backbone
rigidity.  We calculated the number of these critical {\it red} bonds,
$N_R$, and their scaling behavior at the percolation point is also shown
in Fig. 4.  They scale as $N_R \sim L^x$, with $x=0.85 \pm 0.05$. This
is consistent with $x = 1/ \nu$, although we have no analytic argument
for why this should be so.
\\
Previous work on the elastic exponents of regular triangular, central
force, networks have produced conflicting results.  Although the early
work$^5$ gave an exponent in the range, $1.3 \le f/\nu \le 2.0$, later
work suggested that the central force and bond-bending (angular force)
problems were in the same universality class$^{8,15,16}$, so that$^{17}$
$f/\nu \sim 3.0$.  There have even been suggestions$^{15,16}$ that in
2-d, site percolation has exponent near $f/\nu \sim 1.0$, while bond
percolation has exponent near $f/\nu \sim 3.0$. Finally, there is a
recent mean field theory$^4$ which gives exponent $f \sim 1.5$.
\\
The difficulty in obtaining good estimates have been ascribed to:  a)
unusually strong accumulation of roundoff errors$^{7}$, and b) lack of
precision$^8$ in the estimate of $p_c$.  We find that roundoff errors
are largely eliminated if we use the graph theory method to remove all
non-stressed bonds before applying the conjugate gradient method.  In
addition we know $p_c$ exactly for each configuration, so we do not have
to study a range of $p$ using an interative solver.  Thus we have been
able to study the elastic constants for  lattice sizes which were
previously inaccessible (up to linear size $L=512$ - see Fig. 5).
\begin{figure}[] \vbox{ 
\centerline{ \psfig{figure=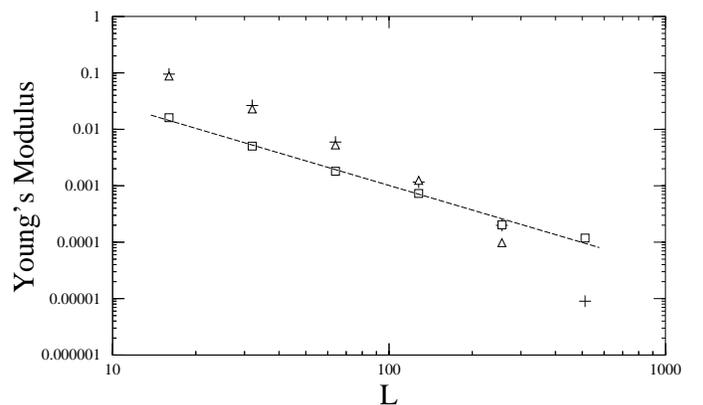,width=9cm,angle=090} }
\centerline{}
\caption{ Finite-size-scaling behavior of the Young's modulus for:
regular lattices (+); lattices randomly displaced by $0.2$ ($\triangle$)
and; lattices with random bond angles ($\Box$).  The lattice sizes used
(number of configurations) were as follows; $16(20,000)$, $32(10,000)$,
$64(1,000-2,000)$, $128(100-300)$, $256(20-40)$, $512(4-6)$.
}
\label{fig:five}
} \end{figure}
As expected, a certain number of the ``generic'' backbones are not rigid
on a regular lattice due to degeneracies. However, the fraction of the
 backbones that are non-rigid on regular lattices increases {\it very
slowly} with lattice size, and is still only $\sim 50\%$ at $L = 512$.
Now note that if the sites of a generic backbone, which is non-rigid on
a regular lattice, are displaced by a small amount $\Delta$, the elastic
 modulus of that backbone is $O(\Delta^2$).  Thus, the elastic constants
 of generic backbones are usually non-universal, for sizes accessable to
simulation, even for lattices displaced by $0.2$ (see Fig. 5).
\\
To avoid the slow size effect caused by proximity to the regular lattice
limit, we also studied a model where the locations of the elements of
the elastic matrix were set by the connectivity of the backbone. To
mimic the highly displaced lattice (large $\Delta$) limit, we assign
each bond an angle to the x-axis which is drawn from a random
distribution of angles (on the interval $[0,360]$, and calculate the
elastic constant using these angles in the force equations.  The results
for this ``random angle model'' are also shown in Fig. 5 (each present
bond has unit spring constant).  We found that the value $f/\nu \sim
1.45$ is quite universal in this limit.
\\
\noindent {\bf Acknowledgements}  
We acknowledge support from PRF, by the DOE under contract
DE-FG02-90ER45418, and by the Humboldt foundation (PD).  We thank Paul
Leath, Bruce Hendrickson, Mike Thorpe and Don Jacobs for useful
discussions.
\newpage
  
\noindent {\bf References}\\
 $^1$ E. Guyon, S. Roux, A. Hansen, D. Bideau, J.-P. Troadec and H.
 Crapo, Rep. Prog. Phys. {\bf 53}, 373 (1990)\\
 $^2$ J.C. Phillips, J. Non-Cryst. Sol. {\bf 43}, 37 (1981); M.F.
 Thorpe, J. Non-Cryst. Sol. {\bf 57}, 355 (1983)\\
 $^3$ M. Rubinstein, L. Leibler and J. Bastide, Phys. Rev. Lett. {\bf
 68}, 405 (1992)\\
 $^4$ S.P. Obukhov, Phys. Rev. Lett. {\bf 74}, 4472 (1995)\\
 $^5$ S. Feng and P.N. Sen, Phys. Rev. Lett. {\bf 52}, 216 (1984)\\
 $^6$ M.A. Lemieux, P. Breton and A.-M.S. Tremblay, J. de Physique {\bf
 46}, L-1 (1985)\\
 $^7$ A.R. Day, R.R. Tremblay and A.-M.S. Tremblay, Phys. Rev. Lett.
 {\bf 56}, 2501 (1986)\\
 $^8$ A. Hansen and S. Roux, Phys. Rev. {\bf B40}, 749 (1989)\\
 $^9$ G. Laman, J. Eng. Math. {\bf 4}, 331 (1970)\\
 $^{10}$ L. Lovasz and Y. Yemini, Siam J. Alg. Disc. Meth. {\bf 3}, 91
 (1982)\\
 $^{11}$ A. Recski, Disc. Math. {\bf 108}, 183 (1992)\\
 $^{12}$ B. Hendrickson, Siam J. Comput. {\bf 21}, 65 (1992); Bruce
 Hendrickson, private communication.\\
 $^{13}$ D. Jacobs and M.F. Thorpe to be published\\
 $^{14}$ M.F. Thorpe and E.J. Garboczi, Phys. Rev. {\bf B35}, 8579
 (1987)\\
 $^{15}$ S. Arbabi and M. Sahimi, Phys. Rev. {\bf B47}, 695 (1993)\\
 $^{16}$ M. Knackstedt and M. Sahimi, J. Stat. Phys. {\bf 69}, 887
 (1992)\\
 $^{17}$ J.G. Zabolitzky, D.J. Bergman and D. Stauffer, J. Stat. Phys.
 {\bf 44}, 211 (1986)\\
 $^{20}$ See for example: C.~H.~Papadimitriou and K.~Steiglitz,
 "Combinatorial Optimization: Algorithms and Complexity", Prentice Hall,
 1982. \\
 $^{21}$ C. Moukarzel, J.~Phs.~A: Math. Gen. {\bf 29} (1996), 8097; (
 physics/9612013). \\
 $^{22}$ C. Moukarzel, P.M. Duxbury and P.L. Leath, to be published.\\
 $^{23}$ C. Moukarzel and P.M. Duxbury to be published.\\
\end{document}